\documentclass[11pt,twoside]{article}

\usepackage{asp2006}
\usepackage{epsf}
\usepackage{psfig}
\usepackage{lscape}
\usepackage{graphicx}

\begin{document}
\title{Observing the molecular composition of galaxies}   
\author{Sergio Mart\'{\i}n}   
\affil{Harvard-Smithsonian Center for Astrophysics}    

\begin{abstract} 
The recent availability of wideband receivers and high sensitivity instruments in the mm and submm wavelengths has opened the possibility of studying in
detail the chemistry of the interstellar medium in extragalactic objects.
Within the central few hundred parsec of galaxies, we find enormous amounts of molecular material fueling a wide variety of highly energetic events
observed in starbursts (galaxies undergoing an intense burst of star formation) and active galactic nuclei (AGN, where activity is driven by the
accretion of material onto the nuclear black hole).

Here it is presented a brief summary of both the history and the latest results in observational chemistry in distant galaxies.
It will be shown how the molecular emission, is a powerful tool to explore the physics of the dust-enshrouded,
buried nuclei of distant ultraluminous galaxies, which are heavily obscured at other wavelengths.
Special attention will be given to the possibilities offered by next generationinstruments such as ALMA (Atacama Large Millimeter Array),
expected to have a vast impact on the field of Extragalactic Chemistry.
Molecular studies in the early Universe will become available at unprecedented sensitivity and resolution.
\end{abstract}


\section{Introduction: A brief look back}

It was back in the mid '30s when the discovery of a series of diffuse unidentified interstellar lines by \citet{Merrill34} aroused the interest
in interstellar molecules in space.
Several species such as CO$_2$, Na$_2$ and NaK were invoked to explain these lines.
It was shortly after, when \citet{Swings37} identified the spectral feature observed by \citet{Dunham37} at $\lambda=4300.3 \, \rm \AA$ as due to the CH molecule.
Being the first accurate identification of a molecular species in space, this detection became the starting point of the fruitful field
of astronomical molecular spectroscopy.
This field has experienced an enormous progress during the last four decades thanks to the development of sensitive instruments operating in radio frequencies,
where most of the rotational molecular spectrum lies.
Almost three decades afer the identification of CH, \citet{Weinreb63} carried out the first molecular radio observation,
namely OH in absorption towards the supernova remnant Cassiopeia A.
It was followed by a series of pivotal detections such as that of NH$_3$ \citep{Cheung68}  and CO \citep{Wilson70},
marking the starting line of astrochemistry.
A few years later, the chemistry in external galaxies would be opened with the CO detection towards M\,82 and NGC\,253 by \citet{Rickard75}.

\begin{table}[!t]
\begin{center}
\caption{Complete census of extragalactic molecular species detected to date. Isotopical substitutions are also included.}
\label{tab:census}
\begin{tabular}{l @{} l @{} l @{\,\,\,} l @{\,\,\,} l @{\,\,\,} l}
\hline
{\bf 2 atoms}	&	{\bf 3 atoms}	&	{\bf 4 atoms}	&	{\bf 5 atoms}	&	{\bf 6 atoms}		&	{\bf 7 atoms}	\\
\hline
\hline
OH 		&	H$_2$O		&	H$_2$CO		&	c-C$_3$H$_2$	&	CH$_3$OH		&	CH$_3$C$_2$H	\\
CO {\tiny \hspace{-5pt} $\bigg\{ \hspace{-5pt} \begin{array}{l} ^{13}CO \\ C^{18}O \\ C^{17}O \\ \end{array} $}
			&	HCN {\tiny \hspace{-5pt} $\bigg\{ \hspace{-5pt} \begin{array}{l} H^{13}CN \\ HC^{15}N \\ DCN \\ \end{array} $}
					&	NH$_3$		&	HC$_3$N		&	CH$_3$CN		&			\\
H$_2$, {\tiny $HD$}	&	HCO$^+$ {\tiny \hspace{-5pt} $\bigg\{ \hspace{-5pt} \begin{array}{l} H^{13}CO^+ \\ HC^{18}O^+ \\ DCO^+ \\ \end{array} $}
					&	HNCO		&	CH$_2$NH	&				&			\\
CH		&	C$_2$H		&	H$_2$CS		&	NH$_2$CN	&				&			\\
CS {\tiny \hspace{-5pt} $\big\{ \hspace{-5pt} \begin{array}{l} ^{13}CS \\ C^{34}S \\ \end{array} $}
			&	HNC {\tiny \hspace{-5pt} $\big\{ \hspace{-5pt} \begin{array}{l} HN^{13}C \\ DNC \\ \end{array} $}
					&	HOCO$^+$	&			&				&			\\
CH$^+$		&	N$_2$H$^+$, {\tiny $N_2D^+$}
					&	C$_3$H		&			&				&			\\
CN						&	OCS								&	H$_3$O$^+$	&			&				&			\\
SO, {\tiny $^{34}SO$}	&	HCO		&			&			&				&			\\
SiO		&	H$_2$S		&			&			&				&			\\
CO$^+$		&	SO$_2$		&			&			&				&			\\
NO		&	HOC$^+$		&			&			&				&			\\
NS		&	C$_2$S		&			&			&				&			\\
LiH		&	H$_3^+$		&			&			&				&			\\
\hline
\end{tabular}
\end{center}
\end{table}

40 years after the first molecular ratio observation, a total of 151 molecular species have been detected in space \footnote{as compiled in astrochemistry.net$<$Oct,2007$>$}.
The vast majority of them have been detected towards the extremely rich chemical environments of hot molecular cores associated to massive star formation
and towards the envelopes of evolved stars.
Outside the Milky Way, on the other hand, beam dilution effects altogether with much broader spectral line features prevent us form such a prolific detection.
As shown in Table~\ref{tab:census}, with the recent addition of the detection of H$_3$O$^+$ towards M\,82 and Arp\,220 \citep{Tak07} the number of molecular species
detected outside the Milky Way has risen to 40, plus 16 isotopical substitutions of these species.
We notice how only some of the simplest complex organic molecules have been detected in the extragalactic interstellar medium (ISM).
Several key species in the organic chemistry such as ethanol ($\rm C_2H_5OH$) and formic acid (HCOOH) have been elusive to several detection attempts.

\section{Extragalactic Interstellar Medium}
Most of the molecular emission observed in galaxies stems from giant molecular clouds complexes (GMCs) of dense ($>10^3\rm cm^{-3}$)
and cold ($\rm T\sim10-50\,K$) gas.
These GMCs with masses of $10^4-10^6\,M_\odot$ and sizes of $30-50$\,parsecs do represent the future birthplaces of stars.
The study of the molecular emission allows us to extract not only the distribution but also the kinematics and physical conditions (density and
temperature) of the molecular gas component in galaxies \citep[e.g. the M\,31 mapping of the CO emission by][]{Nieten06}.

Molecular gas is the fuel powering the energetic events taking place in the nuclear regions of galaxies such as starbursts (SBs), where enhanced star formation
is observed, or active galactic nucleus (AGN), as a result of the accretion onto a supermassive black hole at the center of the galaxy.
The main heating mechanisms, closely related to the nuclear energetic events, affecting the molecular material are: the pervasive and photodissociating UV
radiation of young OB associations; the presence of shock waves due to cloud-cloud collisions, expanding envelopes driven by stellar winds of evolved stars or
surpernovae explosions; X-ray radiation at the vicinity of the nuclear black hole; and cosmic rays.
These heating mechanisms determine the chemistry and therefore the molecular composition of the ISM in galaxies.
While most of the molecular material is in the form of molecular hydrogen ($\gg99\%$), the emission of the different molecular species other than H$_2$,
and thus the study of the chemical composition, provides a key information on the dominating heating mechanisms and physical processes on each galaxy nucleus.

\section{The molecular composition in starbursts}
Due to the brightness of their lines and the prolific molecular detection towards starburst galaxies, these sources have turned into the target of the vast
majority of molecular studies outside the Milky Way.
Among the two most extensively studied examples of starbursts we find NGC\,253 and M\,82.

\begin{figure}
\centering
\includegraphics[angle=-90,width=0.94\textwidth]{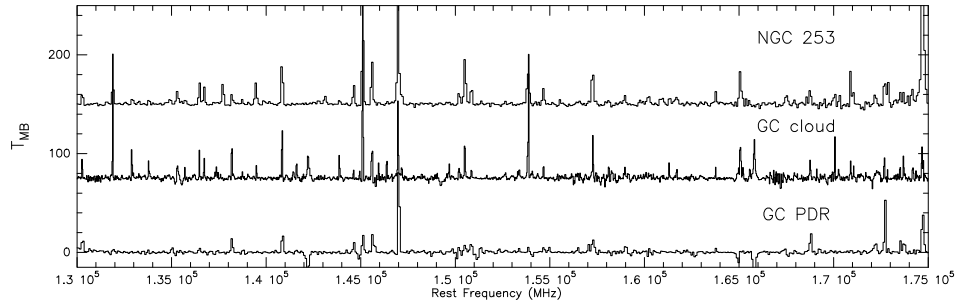}
\caption{{\it Upper spectrum}: Composite emission spectra of NGC\,253 in the 2\,mm atmospheric window \citep{Martin06b}.
{\it Middle and Lower spectra}: Comparison on the same 2\,mm window towards a position in the Sgr\,B2 molecular envelope, as example of a Galactic Center giant
molecular cloud, and a position in the Sgr\,A$^*$ circumnuclear disk (CND), as an example of Galactic photodissociation region (Mart\'{\i}n et al. In prep.).
The spectrum of NGC\,253 clearly resembles that in GC molecular clouds, suggesting a similar origin for their chemistries.}
\label{fig:comparison}
\end{figure}

\begin{figure}
\centering
	\includegraphics[width=0.9\textwidth]{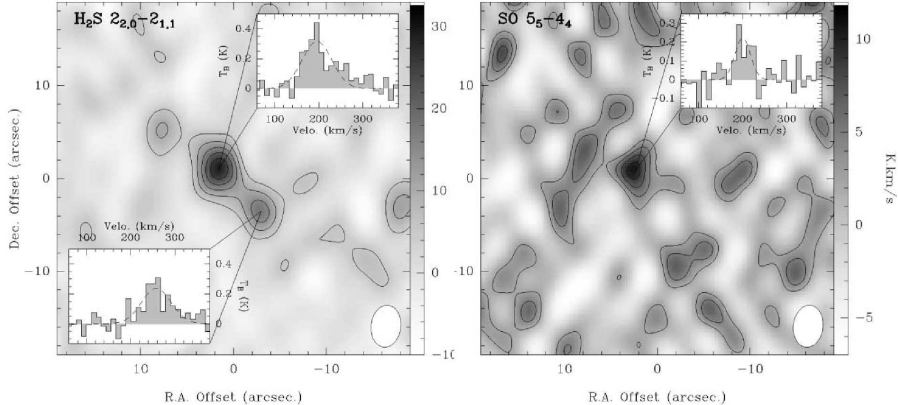}
	\caption{
		High-resolution observations of H$_2$S and SO towards the starburst galaxy NGC\,253 with the Submillimeter Array (SMA).
		The emission from these molecules trace the warm dense molecular material associated with the ongoing star formation within
		the nuclear region of NGC\,253.
		Figure extracted from \citet{Minh07}.
                }
                \label{fig:NGC253}
\end{figure}

{\it NGC\,253} was recently the target of the first unbiased millimeter molecular line survey in the 2\,mm atmospheric window with the IRAM 30m telescope
\citep{Martin06b}. This survey (shown in the upper spectrum of Fig.~\ref{fig:comparison}) not only leaded to the detection of new extragalactic species 
such as SO$_2$, NO and NS \citep{Martin03}, out of the more than a hundred of identified transitions, but it also represented the first step on the spectral
classification of the molecular gas in the nuclei of galaxies.
Such complete studies allowed the description, for the first time, of the chemistry of sulfur-bearing molecules in NGC\,253 as well as the direct comparison
with chemical templates within the galaxy such as that shown in Fig.~\ref{fig:comparison}.
NGC\,253 shows a molecular emission significantly comparable to that observed towards the GMCs we find in the Galactic Center region.
This comparison suggests that similarly to what we observe towards GMCs, the molecular gas in NGC\,253 is mainly heated by low-velocity shock waves,
likely due to cloud-cloud collisions in a barred potential orbital motions.
High resolution imaging with instruments such as the SMA, allows us to trace the warm and dense molecular cloud complexes within the nuclear region of NGC\,253,
which represent the fuel of the intense observed star formation in this galaxy.
Fig~\ref{fig:NGC253} shows an example of the observation of two sulfur-bearing species such as H$_2$S or SO \citep{Minh07}.

{\it M\,82}, on the other hand shows a particularly interesting chemistry. It was known since long the surprisingly relative low abundances derived by
the non detection of several species such as SiO, HNCO and CH$_3$OH \citep{Henkel87,Mauers91,Nguyen91} as well as the non detection of a hot NH$_3$ component
that was observed in other starbursts \citep{Weiss01,Takano02,Mauers03}.
The recent detection of widespread emission of species commonly used as photodissociation tracers, namely HCO, CO$^+$, HOC$^+$, and C$_2$H
\citep{Burillo02,Fuente06}, turned the nuclear region of M\,82 into the prototype of extragalactic photodissociation region (PDR).
Nevertheless, the nuclear region of galaxies is an heterogeneous composite of cloud complexes affected by different environmental conditions.
It is therefore difficult to find pure chemical prototypes in external galaxies.
This is particularly suggested by the detection of methanol in M\,82 with abundances
well above that expected by pure gas-phase chemistry, which is an evidence of the presence of a significant amount of molecular material
shielded from the pervading UV radiation and similar to that found in other starbursts galaxies such as NGC\,253 or Maffei\,2 \citep{Martin06a}.

{\it NGC\,4945} has been also the target of one of the most ambitious multi-line studies of the molecular content of the ISM in a starburst galaxy.
A total of 80 transitions of 19 species were observed with SEST 15\,m telescope in the range between 82 and 354\,GHz towards this galaxy \citep{Wang04}.

The overall comparison of the fractional abundances observed towards these prototype galaxies has allowed to build up a picture
of the chemical evolutionary sequence of their nuclear starbursts.
Within this sequence, NGC\,253 and M\,82 would represent the extreme evolutionary states.
Thus NGC\,253 appears to be in an early stage with huge amounts of dense molecular gas feeding the ongoing burst of star formation where the dominant
chemical fingerprints we observe are those of the shocks affecting this material.
On the other side, M\,82, with a similar star forming rate, has consumed most of its molecular gas reservoir and its ISM shows the traces of the
photodissociation produced by the young hot stars formed during star burst period.
Galaxies such as NGC\,4945 would represent a prototype of intermediate stage between those of NGC\,253 and M\,82 \citep{Wang04,Martin06b}.

\begin{figure}
\centering
\includegraphics[width=0.91\textwidth]{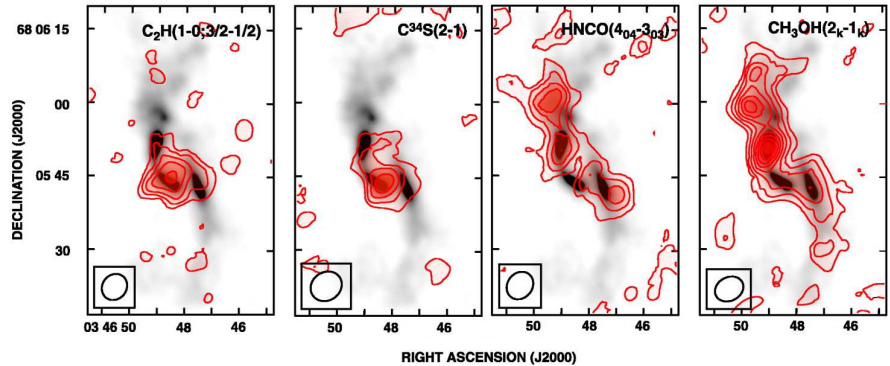}
\caption{Integrated intensity of $\rm C_2H,\,C^{34}S,\,HNCO,\,and\,CH_3OH$ towards the nuclear region of IC\,342 with the $^{12}$CO (1-0) in grey scale 
for comparison. Figure extracted from \citet{Meier05}.}
\label{fig:IC342}
\end{figure}

The understanding of the chemical complexity in galaxies altogether with the high spatial resolution provided by current interferometers enables us
to spatially disentangle the different physical processes taking place in the nuclear region of galaxies.
That is the case of the high angular resolution molecular line observations towards IC\,342 carried out by \citet{Meier05}.
The mapping of selected molecular tracers (Fig.~\ref{fig:IC342})
show how the molecular gas is highly pervaded by the photodissociation radiation around the nuclear star clusters (traced by $\rm C_2H,\,and\,C^{34}S,$)
while the denser and warmer material is observed towards the circumnuclear spiral arms where the gas is mostly affected by cloud-cloud collisions (traced
by HNCO and CH$_3$OH), in agreement with the SiO observations by \citet{Usero06}.

\section{Beyond starbursts}

\begin{figure}
\centering
\includegraphics[width=0.9\textwidth]{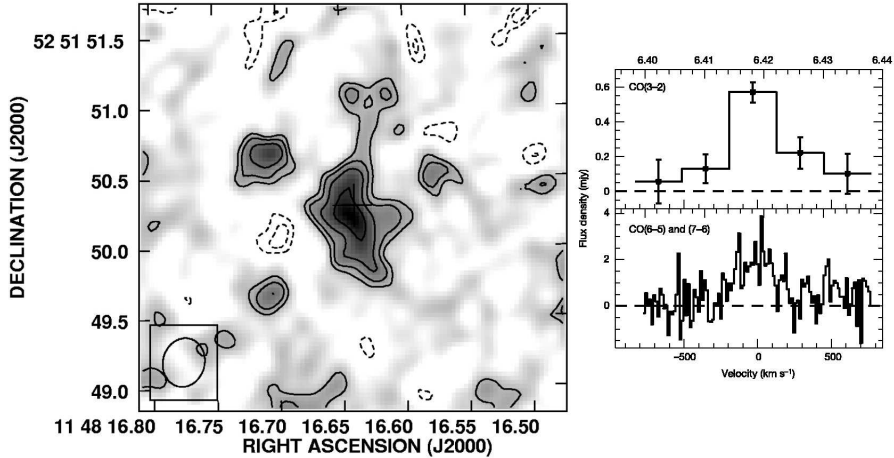}
\caption{
CO observations towards J$1148+5251$, the highest redshift quasar currently known, at a redshift z=6.42
which correspond to an age of the Universe younger than 1\,Myr.
{\it Left:} CO (3-2) obtained with the VLA. A total of $\sim60$\,hours of integration time were necessary to get the resolved CO structure of this source
{\it Right:} CO transitions detected with the VLA ({\it upper}) and PdBI ({\it lower}).
Figures extracted from \citet{Walter03,Walter04}.
}
\label{fig:J1148}
\end{figure}

During the last few years, an increasing interest is coming up on unveiling the molecular gas content of galaxies at early times as this is likely to be
directly related to the star formation history of the Universe.
As shown in Fig.~\ref{fig:J1148}, molecular gas has been detected up to a redshift of z=6.42 \citep{Walter03},
when the Universe had an age below 1 million years.
Unfortunately, we are still far from being able to study the chemical composition of these sources in such detail as we reach in nearby galaxies.
The limited sensitivity of current instruments prevent us from going further in the search for molecules more complex than CO as evidenced
by the recent non detection of HCN towards J$1148+5251$ \citep{Riechers07}.
This is the reason of the increased effort to study the chemistry of Ultra Luminous Infrared Galaxies (ULIRGs) which,
with enormous star formation rates, are thought to be reasonable templates of the ISM conditions at high redshift.
Among these studies, Arp\,220 (at a z$\sim0.02$) has been the target of recent sensitive multiline studies \citep{Greve06}.

At not so high redshifts, when the Universe was $\sim2-4$\,Myr old, detections have been obtained of HCN and HCO$^+$.
These observations, close to the limit of current instrumentation capabilities, require many hours of observing time in order to reach a significant 
detection \citep[e.g.][]{Wagg05,Riechers06}.

Nevertheless, the observation of the molecular material at high redshift provide us with some interesting advantages.
Such is the possibility of detecting the high-J CO transitions as shown by the $J=11-10$ observations towards a z=3.9 quasar by
\citet{Weiss07}. The rest frequency of this transition, at 1.3\,GHz, is almost impossible to be observed from ground telescopes, but it is easily detected with the 
available mm telescopes at its redshifted frequency of $\sim 1$\,mm.

\section{A bright future with ALMA}

The Atacama Large Millimeter Array (ALMA) is creating great expectations among the astronomical comunity.
In particular, ALMA will be a breakthrough in the field of extragalactic chemistry because of two main reasons:
First, the combination of resolution and sensitivity. With baselines ranging from 150\,m to 18\,km, ALMA will reach resolutions between $0.7''$ and
$0.01''$ at 1\,mm with a sensitivity of more than an order of magnitude with respect of current available mm and sub-mm instruments.
It will be possible to observe the nearby galaxies with the resolution we have currently in the Galactic Center with single dish telescopes and we will 
be able to fully resolve the molecular emission of galaxies at the early ages of the Universe.
Similar to what we observe today in the Galaxy, we will be able to study in detail the chemical complexity towards the star forming region in nearby galaxies;
Secondly, the instantaneous wideband of at least 8\,GHz will allow the mapping of the emission of many molecular lines at a time.
With an unprecedented observing speed, as compared to the hundreds of hours which would require nowadays, ALMA will provide us with similar line surveys
as those in Fig.~\ref{fig:comparison} with the additional spatial distribution of each of the molecular species.

ALMA (the Spanish word for {\it SOUL}) will be the starting point of the study of the complex organic molecules in external galaxies.
This will open the possibility of observing the emission of complex bio-molecules from nearest to the highest redshifted galaxies
and therefore we will get a hint on the evolution of the building blocks of life through the history of the Universe.





\begin{thebibliography}{}
\bibitem[Cheung et al.(1968)]{Cheung68} Cheung, A.~C., Rank, D.~M., Townes, C.~H., Thornton, D.~D., \& Welch, W.~J.\ 1968, Physical Review Letters, 21, 1701 
\bibitem[Dunham(1937)]{Dunham37} Dunham, T., Jr.\ 1937, \pasp, 49, 26
\bibitem[Fuente et al.(2006)]{Fuente06} Fuente, A., Garc\'{\i}a-Burillo, S., Gerin, M. et al. 2006, \apj, 641, L105
\bibitem[Garc\'{\i}a-Burillo et al.(2002)]{Burillo02} Garc\'{\i}a-Burillo, S., Mart\'{\i}n-Pintado, J., Fuente, A., \& Usero, A. 2002, ApJ, 575, L55
\bibitem[Greve et al.(2006)]{Greve06} Greve, T.~R., Papadopoulos, P.~P., Gao, Y., \& Radford, S.~J.~E.\ 2006, ArXiv Astrophysics e-prints, arXiv:astro-ph/0610378 
\bibitem[Henkel et al.(1987)]{Henkel87} Henkel, C., Jacq, T., Mauersberger, R., Menten, K. M., \& Steppe, H. 1987, \aap, 188, L1
\bibitem[Mart{\'{\i}}n et al.(2003)]{Martin03} Mart{\'{\i}}n, S., Mauersberger, R., Mart{\'{\i}}n-Pintado, J., Garc{\'{\i}}a-Burillo, S., \& Henkel, C.\ 2003, \aap, 411, L465 
\bibitem[Mart{\'{\i}}n et al.(2005)]{Martin05} Mart{\'{\i}}n, S., Mart{\'{\i}}n-Pintado, J., Mauersberger, R., Henkel, C., \& Garc{\'{\i}}a-Burillo, S.\ 2005, \apj, 620, 210 
\bibitem[Mart{\'{\i}}n et al.(2006a)]{Martin06a} Mart{\'{\i}}n, S., Mart{\'{\i}}n-Pintado, J., \& Mauersberger, R.\ 2006a, \aap, 450, L13 
\bibitem[Mart{\'{\i}}n et al.(2006b)]{Martin06b} Mart{\'{\i}}n, S., Mauersberger, R., Mart{\'{\i}}n-Pintado, J., Henkel, C., \& Garc{\'{\i}}a-Burillo, S.\ 2006b, \apjs, 164, 450 
\bibitem[Mauersberger \& Henkel(1991)]{Mauers91} Mauersberger, R., \& Henkel, C. 1991, \aap, 245, 457
\bibitem[Mauersberger et al.(2003)]{Mauers03} Mauersberger, R., Henkel, C., Wei{\ss}, A., Peck, A.~B., \& Hagiwara, Y.\ 2003, \aap, 403, 561 
\bibitem[Meier \& Turner(2005)]{Meier05} Meier, D.~S., \& Turner, J.~L.\ 2005, \apj, 618, 259 
\bibitem[Merrill(1934)]{Merrill34} Merrill, P.~W.\ 1934, \pasp, 46, 206 
\bibitem[Minh et al.(2007)]{Minh07} Minh, Y.~C., Muller, S., Liu, S.-Y., \& Yoon, T.~S.\ 2007, \apjl, 661, L135
\bibitem[Nguyen-Q-Rieu et al.(1991)]{Nguyen91} Nguyen-Q-Rieu, Henkel, C., Jackson, J. M., \& Mauersberger, R. 1991, \aap, 241, L33
\bibitem[Nieten et al.(2006)]{Nieten06} Nieten, C., Neininger, N., Gu{\'e}lin, M., Ungerechts, H., Lucas, R., Berkhuijsen, E.~M., Beck, R., \& Wielebinski, R.\ 2006, \aap, 453, 459 
\bibitem[Rickard et al.(1975)]{Rickard75} Rickard, L.~J., Palmer, P., Morris, M., Turner, B.~E., \& Zuckerman, B.\ 1975, \apjl, 199, L75 
\bibitem[Riechers et al.(2006)]{Riechers06} Riechers, D.~A., Walter, F., Carilli, C.~L., Weiss, A., Bertoldi, F., Menten, K.~M., Knudsen, K.~K., \& Cox, P.\ 2006, \apjl, 645, L13 
\bibitem[Riechers et al.(2007)]{Riechers07} Riechers, D.~A., Walter, F., Carilli, C.~L., \& Bertoldi, F.\ 2007, ArXiv e-prints, 710, arXiv:0710.4525 
\bibitem[Swings \& Rosenfeld(1937)]{Swings37} Swings, P., \& Rosenfeld, L.\ 1937, \apj, 86, 483
\bibitem[Takano et al.(2002)]{Takano02} Takano, S., Nakai, N., \& Kawaguchi, K.\ 2002, \pasj, 54, 195 
\bibitem[Usero et al.(2006)]{Usero06} Usero, A., Garc{\'{\i}}a-Burillo, S., Mart{\'{\i}}n-Pintado, J., Fuente, A., \& Neri, R.\ 2006, \aap, 448, 457
\bibitem[Van der Tak et al.(2007)]{Tak07} Van der Tak, F., Aalto, S., \& Meijerink, A\&A, submitted (2007).
\bibitem[Wagg et al.(2005)]{Wagg05} Wagg, J., Wilner, D.~J., Neri, R., Downes, D., \& Wiklind, T.\ 2005, \apjl, 634, L13 
\bibitem[Walter et al.(2003)]{Walter03} Walter, F., et al.\ 2003, \nat, 424, 406 
\bibitem[Walter et al.(2004)]{Walter04} Walter, F., Carilli, C., Bertoldi, F., Menten, K., Cox, P., Lo, K.~Y., Fan, X., \& Strauss, M.~A.\ 2004, \apjl, 615, L17 
\bibitem[Wang et al.(2004)]{Wang04} Wang, M., Henkel, C., Chin, Y.-N., Whiteoak, J.~B., Hunt Cunningham, M., Mauersberger, R., \& Muders, D.\ 2004, \aap, 422, 883 
\bibitem[Weinreb et al.(1963)]{Weinreb63} Weinreb, S., Barrett, A.~H., Meeks, M.~L., \& Henry, J.~C.\ 1963, \nat, 200, 829 
\bibitem[Wei{\ss} et al.(2001)]{Weiss01} Wei{\ss}, A., Neininger, N., Henkel, C., Stutzki, J., \& Klein, U.\ 2001, \apjl, 554, L143 
\bibitem[Wei{\ss} et al.(2007)]{Weiss07} Wei{\ss}, A., Downes, D., Neri, R., Walter, F., Henkel, C., Wilner, D.~J., Wagg, J., \& Wiklind, T.\ 2007, \aap, 467, 955
\bibitem[Wilson et al.(1970)]{Wilson70} Wilson, R.~W., Jefferts, K.~B., \& Penzias, A.~A.\ 1970, \apjl, 161, L43 
\end{thebibliography}
\end{document}